# Personal Green IT Use: Findings from a Literature Review

## Full research paper

### Ayodhya Wathuge
Faculty of Business, Law, and Arts
Southern Cross University
Bilinga, Queensland
Email: ayodhyawathuge@gmail.com

### Darshana Sedera
Faculty of Business, Law, and Arts
Southern Cross University
Bilinga, Queensland
Email: darshana.sedera@scu.edu.au

### Golam Sorwar
Faculty of Science and Engineering
Southern Cross University
Bilinga, Queensland
Email: golam.sorwar@scu.edu.au

## Abstract

Research addressing the greening of internet user behaviours at hedonic and utilitarian levels is scarce. To identify dimensions, scales and strong relationships arising from motivation, we reviewed a sample of research articles related to the personal green IT context. We used Self-determination theory as the theoretical framework to categorize factors into different motivation dimensions. A qualitative literature review analyses five pair-wise associations between motivation constructs of the theory and green IT use. This work builds on the prior research related to environmental motivation by summarizing the measures applied to the evaluation of personal green IT behaviours and by examining the relationships broadly defined in the Self-determination theory, distinguishing between hedonic and utilitarian green IT use.

**Keywords** personal green IT use, hedonic, utilitarian, motivation, self-determination theory.

# 1 Introduction

It is less known that each Gigabyte of internet use leaves a carbon dioxide ($CO_2$) footprint of up to 63 g and uses up to 35 Litres of clean water (Obringer et al., 2021). Such a large amount of environmental pollution resulting from the internet is mainly attributed to the operations of data centres – the internet infrastructure (Han et al., 2019). To mitigate the internet's impact on the environment, many attempts have been made at the global, organizational, and individual levels, involving both technical and non-technical endeavours (UN, 2019, UNFCCC, 1992). However, the individual's role in using internet services responsibly in one's personal circumstances (henceforth referred to as responsible personal internet use[1]) has rarely received much attention (Hill et al., 2020, Sedera et al. 2017; Seidler et al., 2020). While there are studies on the individual level, most of such work addresses an individual in the organizational setting. The emphasis on personal internet use is critically important to climate change discussions, considering the estimated 414 Kg of $CO_2$ emitted annually by an individual using digital devices (Griffiths, 2020).

In order to build a sustainable digital world where new digital technologies are continuously introduced (Sedera and Lokuge 2017), it is also essential to identify the personal consumption level actions that can be taken to minimize the adverse environmental effects of the internet (Deng et al., 2017). The internet is a vital tool for human life, and its use can be conceived as either *hedonic* or *utilitarian* (Lowry et al. 2012; To et al. 2007; Van der Heijden 2004). Hedonic internet use, similar to past studies, refers to when individuals use the internet to gain the experiential value of joy or fun. For example, an individual would use social media like Facebook for the pleasure they get, while another individual uses the internet to do online shopping. Accordingly, both above activities are hedonic in nature. On the other hand, utilitarian internet use refers to when individuals use the internet to gain instrumental value. Such actions could be performing professional work such as attending one's job-related video conferences and accessing services like e-banking or e-learning using their personal internet infrastructure. It is essential to distinguish the differences between the two types of internet use, as the strategies require employing in reducing the emissions of each class could be different (Grant et al. 2007). For example, Turel (2015) shows that quitting the addiction to hedonic information systems like Facebook requires motivations related to personal norms such as guilt, while Brivio et al. (2018) show that to reduce the technostress that causes due to the usage of technology for utilitarian purposes, external motivations such as well-designed technologies that suit different age groups are required (Sedera and Lokuge 2020; Sedera et al. 2016). Acknowledging the importance of two types and highlighting of different motivation pathways each category requires, we identify internet use in two categories: personal level hedonic and utilitarian.

In order to identify strategies that can be taken to instil responsible personal internet usage, we looked at personal level green (environmentally friendly) information technology (IT) behaviours such as the adoption of smart meters to track domestic electricity usage and usage of game-based techniques to enhance environmental conservation (Ke et al., 2019, Loock et al., 2013; Wathuge and Sedera 2021). At the personal level, they have identified that psychological and external factors such as monetary and non-monetary incentives/disincentives motivate individuals to engage in green IT behaviour (Panzone et al. 2020; Wathuge and Sedera 2022; Wunderlich et al. 2019). While past researchers have identified that *motivation* plays a significant role in predicting green IT behaviour of individuals (Koo et al. 2015; Molla 2009), the current research lacks comprehensive knowledge on (i) types of motivations and their impact on personal green IT behaviour and (ii) a distinction between how the types of motivations impact the hedonic and utilitarian green IT behaviours. We conducted a literature review on personal level green IT behaviours deriving the above findings. The study aimed to determine the motivation factors that affect green internet user behaviours at personal hedonic and utilitarian levels. In doing so, we derived the motives aligning with self-determination theory (SDT), a heavily used theory to depict individual level pro-environmental behaviours (Pelletier et al., 1998). The study's objective is to capture the wealth of research linking motivation to personal level green IT behaviours at hedonic and utilitarian aspects. In doing so, the paper provides opportunities for researchers as to where one should focus on greening one's hedonic and utilitarian internet use.

The section that follows investigates various competing models of green IT behaviour and explains why the SDT was chosen as the organizing framework for this research analysis. An explanation of the procedures used to collect and categorize the research with a more extensive discussion of the literature review is also provided. The findings of this literature review are then provided in accordance with the 24 papers that were reviewed. The study is organized around the SDT components and the links between the constructs in the context of green IT. It explains where the current research stands and where future research is required. Finally, the paper ends with a conclusion.

---

[1] Responsible internet use – Using the internet with a deep understanding of the importance to take responsibility for environmental damage (including how one's online behaviour and activity affect the environment) (Lewin et al. 2021).

## 2 Theoretical Perspective

Researchers have derived several theories to explain personal level green IT behaviours. A few examples are the unified theory of acceptance and use of technology (UTAUT) (Brauer et al., 2016), theory of reasoned action (TRA) (Chen et al., 2016), motivation theory (Chen et al., 2016, Wunderlich et al., 2013b), theory of planned behaviour (TPB) (Kranz and Picot, 2012), goal framing theory (GFT) (Ke et al., 2019) and technology acceptance model (TAM) (Kranz and Picot, 2012). They all provide different aspects of the psychology of the human mind, depicting how a person can be intrigued to perform an activity. However, self-determination theory offers an individual's perspective and five different taxonomies of regulations (Ryan and Deci 2008). It provides a wide range of capabilities to identify reasons for individual pro-environmental behaviour change concerning IT.

Researchers in environmental sustainability have proved that the five sub-constructs of SDT: intrinsic, integrated, identified, introjected and external are determinants of pro-environmental behaviour which exert a direct impact on an individual's behaviour (Koo et al., 2013, Koo et al., 2015). Here, we exclude amotivation construct, which refers to the absence of motivation because, according to the theory, amotivation does not lead to behaviours. Accordingly, the regulations can be divided based on their intrinsic and extrinsic nature. SDT suggests that in *intrinsic motivation*, a person may manifest an environmentally responsible behaviour due to reasons inherent to the behaviour, such as pleasure and satisfaction (Ryan and Deci 2008). On the other hand, *extrinsic motivation* suggests engaging in behaviour due to reasons external to the activity itself (Pelletier, 2002). Extrinsic motivation is a collection of four motivations that differ according to the level of autonomy. *External regulation* occurs when the person does not internalize the motivations but depends on external societal factors such as rewards or punishments, materialistic benefits such as money and incentives/disincentives, and coercive factors such as rules and regulations (Koo et al. 2015; Ryan and Deci 2008). The second construct of extrinsic motivation, *introjected regulation*, occurs when reasons for engaging in behaviour are endorsed by the person (Ryan and Deci, 2008). However, it is controlled by internal pressures such as obligation and guilt. The third construct of extrinsic motivation is *identified regulation*, where an individual performs a behaviour by valuing it and identifying the reasons for performing it (Ryan and Deci, 2008). The last construct of extrinsic motivation is *integrated regulation* which occurs as people evaluate the value of engaging in a behaviour and bringing it into coherence with other aspects of the self (Ryan and Deci, 2008). We use these five constructs of the SDT theory to identify the motives of personal green IT behaviours.

## 3 Methodology

A qualitative literature review is one of the most well-established methodologies for integrating research findings and assessing the domain's cumulative knowledge (Oliver 1987; Onwuegbuzie et al. 2012). This method enables a researcher to review and assess quantitative and qualitative literature within a discipline in order to get a conclusion on the state of the field (Petter et al., 2008). A qualitative literature review is better suited to explain how the relationships in the literature have been explored; whether there appears to be support for a causal relationship between two variables; and whether there are any potential boundary requirements for the model.

### 3.1 Scope of the review

In retrieving relevant academic research articles for our study, we began the search process by searching the Association of Information System's Senior Scholars' Basket of Journals, and some other peer-reviewed journals such as Computers in Human Behaviour (CHB), Journal of Cleaner Production (JCP), Sustainability (S) and Information Technology and People (ITP). Moreover, we referred to conference proceedings, specifically those of the Americas Conference on Information Systems (AMCIS), and the International Conference on Information Systems (ICIS) as they are the most prestigious conferences in the IS domain. Moreover, we used databases such as Web of Science and Google Scholar to gather any missing academic publications in reputed journals. We limited our search only to full research articles that are in English. To derive related articles, first, we defined the search term aligning with (Dao et al., 2011). Our search terms were (green OR environ* OR sustainab* OR ecolog*) AND (information system OR IS OR IT OR technolog* OR digital OR comput*) AND (behavio* OR intention* OR use OR intervention) AND (motivat* OR intrinsic OR integrated OR identified OR introjected OR external OR pleasure OR norms OR incentives). Simple modifications were made to the key terms to match the respective journal repository requirements. We removed key terms such as 'IS' and 'IT' where required when false matches were rendered. We derived articles published within the last ten years (2012-2022) for the review.

We screened for relevance by scanning the content and analysing the titles and abstracts. If the paper's fit with the criteria was unclear after this screening, we read the full article to decide whether the paper would be included in the final sample. The criteria for inclusion or exclusion of a paper were as follows. First, we checked for the paper's green IT/IS, green computing, green management, or sustainability aspects. Secondly, we checked if the paper addresses a green IT consumerism aspect, such as green IT usage. Thirdly, we checked if the study was conducted at the personal level, which includes people from the public. We excluded publications that are solely focused on employees which relate to an organizational setting. Moreover, we checked if the studies discuss any motivational aspect of the SDT. The studies that did not fit the suggested criterion and did not explore the green IT/IS strategies empirically, despite containing keywords in their abstracts or subject headings, were excluded (Dybå and Dingsøyr 2008). At the end of the review, we derived 24 articles. The low number of academic publications can be attributed to the scarcity of personal level green IT behavioural research on aspects of motivation. Furthermore, the rigorous scanning process excluded some related articles that fall into non-credible academic sources.

## 3.2  Organizing the literature review

The findings from empirical studies on green IT are organized according to the SDT constructs. Subsections are grouped based on the expected causal relationships between paired SDT constructs and green IT behaviour. Collecting the literature in this manner helps to examine whether there is support for each proposed relationship within the SDT. First, we categorized the factors that impact the personal level green IT behaviours into the five constructs of the SDT. Next, we identified the influence of those motivation factors on the dependent variable, attitude towards/intention of/behaviour change, depicted through green use of IT from hedonic or utilitarian perspectives. Green IT behaviours such as using mobile applications, CDs and DVDs and playing online games to curb emissions that incorporated experiential value were categorized into hedonic behaviours. In contrast, green IT behaviour such as using smart electricity meters was categorized as utilitarian behaviours as they provide information for performing a green behaviour. This categorization allows for identifying if there are differences in the strengths of the hedonic or utilitarian use-based relationships.

The five constructs of SDT can be divided based on motivation and perceived locus of control. We identified studies that have used all the five regulations (intrinsic, integrated, identified, introjected and external), regulations based on motivation (intrinsic and extrinsic), or regulations based on perceived locus of control (autonomous and controlled). When categorizing such constructs into the five regulations, we first looked at whether there are question items related to the respective regulations. If there were question items representing different regulation types, we ticked off all the constructs that represented them. Each construct of the SDT has multiple operationalizations; the support, or lack of support, for relationships between constructs may be due to how the constructs were measured. We captured such instances too.

# 4  Results

The results of the review are arranged as follows. In identifying the factors that impact personal green IT behaviours based on five SDT constructs, we looked at different dimensions, measurement items and scales used for each construct. Then we investigated the influence of each SDT construct on green IT use.

## 4.1  Measuring the five motivation constructs

There are different dimensions as well as scales that measure the constructs of the SDT occurring within the papers chosen for this review. This section identifies some of the different operationalizations of each construct. In measuring the constructs, studies have used the perspective of the motivation in conserving the environment (Koo et al. 2013; Koo et al. 2015; Wati and Koo 2012) as well as the perspective of the motivation in engaging in a green IT activity (Wunderlich et al. 2013b; Wunderlich et al. 2012). However, the majority of the studies have incorporated the environmental perspective. The following section reviews the dimensions and scales used to measure the motivation constructs.

In measuring intrinsic motivation, studies have used various dimensions such as perceived pleasurability (Bukchin and Kerret 2020; Wati and Koo 2012), perceived enjoyment (Koo et al., 2015), internal locus of control in engaging in a green IT behaviour (Wunderlich et al. 2013a; Wunderlich et al. 2013b; Wunderlich et al. 2012) and cognitive absorption (Shevchuk et al., 2019). Mainly, items were derived from scales developed by Ryan and Connell (1989), Agarwal and Karahanna (2000) and Lowry et al. (2012). In measuring integrated regulation, studies have used dimensions such as perceived usefulness (Wati and Koo, 2012) and perceived altruism (Wati and Koo, 2012). Items related to perceived usefulness were mainly derived from Taylor and Todd (1995). To measure the identified regulation, studies have used dimensions such as perceived importance (Wati and Koo, 2012). Moreover, studies have measured the construct using items related to the internal locus of control (Wunderlich et al. 2013b; Wunderlich et al. 2012), which were mainly derived from the scales developed by Ryan and Connell (1989). In measuring the introjected regulation, past studies have used dimensions such as ego-involvement (Wati and Koo, 2012) and environmental concern of an individual (Zhang et al., 2020) by modifying scales developed by Lee et al. (2014). Furthermore, measuring personal norms derived from the scales developed by Schwartz (1977) and Steg et al. (2005) could also be identified. Past researchers have used various dimensions to measure the external regulation construct. The main dimensions were societal norms, monetary aspects, and regulatory compliances. Moreover, measuring the external rewards an individual receives, Koo and Chung (2014) have measured external regulation using the perceived usefulness dimension incorporating the perceived usefulness scale developed by Taylor and Todd (1995). Moreover, dimensions such as personal motivation were used to refer to the financial aspect of the construct (Bukchin and Kerret, 2020). Such studies have used Gifford and Nilsson's (2014) scale. Additionally, the perceived cost of technologies was captured through scales developed by Nesbitt (2001). Furthermore, past studies measure external regulation using measures related to normative beliefs such as influences from sources like media (Kranz and Picot, 2012), using dimensions such as interactivity, cooperation, and competition. Moreover, in measuring the perceived locus of control, studies such as (Wunderlich et al., 2019) have used scales developed by Ryan and Connell (1989).

Apart from the above dimensions and scales, most studies have used the central constructs in the SDT theory while using The Motivation Toward the Environment Scale (MTES) developed by Pelletier et al. (1998). MTE Scale has been applied and tested in different contexts of environmental studies across geographical territories (Pelletier et al., 1998, Villacorta et al., 2003), and has demonstrated acceptable levels of reliability and strong convergent and discriminant validity. It provides 18 measures representing all six motivation constructs, including amotivation. What makes this particular instrument to measure environmental motivation unique is this instrument captures the multidimensional and complex nature of motivation by measuring all six constructs and using at least four measures for each dimension.

### 4.2   Pair-wise comparison for motivation constructs

This literature review includes studies on a variety of IT user behaviours. The only prerequisites for inclusion in this review of the literature are that the article (a) presents an empirical result (either quantitative or qualitative) and (b) investigates a relationship broadly defined within the SDT. The following section reviews the influence of each SDT construct and respective green IT behaviours. First, we looked at the utilitarian aspect of green IT use. Towards the end of the section, a consolidated explanation is provided on hedonic use.

#### 4.2.1   Intrinsic → green IT use

There is moderate support for this relationship at the individual utilitarian level analysis within the literature. Koo et al. (2015) show a statistically significant influence of perceived enjoyment on intention to use smart green IT equipment to measure electricity usage in individual IT devices. However, in another study, Koo and Chung (2014) show that intrinsic motivation had a non-significant impact on attitude to use smart green IT to save electricity consumption, which further led to a non-significant effect on the intention to use green IT. Bukchin and Kerret (2020) show that pleasure in engaging in environmental behaviours leads to green IT behaviours. Wunderlich et al. (2012), (Wunderlich et al., 2013a) and (Wunderlich et al., 2013b) show that internal perceived locus of causality (PLOC) had a positive influence on the intention to continue using smart metering IS. These studies demonstrated moderate support for the relationship.

#### 4.2.2   Integrated → green IT use

We could not conclude the influence of the integrated regulation construct on green IT use with only three studies. Adding to the confusion, the study by Koo and Chung (2014) shows that integrated motivation had a non-significant impact on the attitude to use smart green IT to save electricity consumption, which further led to a non-significant effect on the intention to use green IT. However, Bukchin and Kerret (2020) show that considering engaging in environmental behaviours as an integral part of self leads to green IT behaviours. Since the identified influences are opposite in nature, we refrained from concluding on the relationship.

### 4.2.3 Identified → green IT use

At the individual utilitarian level, there is strong support for the influence of identified regulation on green IT use. Koo and Chung (2014) show that identified motivation had a significant impact on attitude to use smart green IT to save electricity consumption. Wati and Koo (2012) showed that identified regulation had a positive influence on intention to use green information systems (IS). Testing on another dimension of the construct, Wunderlich et al. (2012) and Wunderlich et al., (2013a) show that internal PLOC had a positive influence on using smart metering IS. Since all the studies depicted positive results, we concluded that the constructs strongly support the relationship.

### 4.2.4 Introjected → green IT use

There is mixed support for this relationship in the personal utilitarian use analysis within the literature. Koo and Chung (2014) show that introjected motivation had a non-significant impact on attitude to use smart green IT to save electricity consumption, which further created a non-significant effect on the intention to use green IT. However, using another variable, social influence, Koo and Chung (2014) show a positive and statistically significant relationship between the variables (introjected regulation and attitude towards green IS use behaviour leading to intention to use green IS). Bukchin and Kerret (2020) show that personal norms such as guilt led to the intent of engaging in green IT behaviours. Moreover, Wati and Koo (2012) show that introjected regulation did not significantly influence green IT use. On the other hand, Wunderlich et al. (2013b) show that introjected regulation negatively impacted green IT behaviour of electricity consumers. As most studies did not showcase positive or mixed results, we concluded that the constructs have mixed support for the relationship.

### 4.2.5 External → green IT use

There is strong support for this relationship in the personal utilitarian use analysis within the literature. Koo et al. (2015) show a statistically significant influence by extrinsic motivation (saving money and legislative pressure) on intention to use smart green IT equipment to measure electricity usage in an individual's IT appliances. Moreover, Bukchin and Kerret (2020) show that personal motivations such as monetary and materialistic benefits positively correlated to the intention to engage in green IT behaviours. Chen et al. (2017) show a non-significant influence by perceived cost on the intention to adopt green IS. They illustrate that perceived cost did not play a significant role because some utility companies provide free smart meters or waive the installation fee for customers. Koo and Chung (2014) demonstrated that external motivation (perceived usefulness) had a significant impact on the attitude to use smart green IT to save electricity consumption, which further led to creating a non-significant effect on the intention to use green IT. Loock et al. (2013) exhibit the impact of medium-level goal setting through web-based systems in changing household energy use. They explain that goals communicate normative information to the individual by suggesting the level of performance they can expect to attain. Such normative pressure induced by the goal motivates forced individuals to save electricity. Bukchin and Kerret (2020) show a positive influence by societal criticisms an individual receives on the intention to engage in green IT behaviours. Wunderlich et al. (2012), Wunderlich et al. (2013a) and Wunderlich et al. (2013b) demonstrate that external PLOC had a positive influence on the intention to continue using smart metering IS. Kranz and Picot (2012) exhibit a positive influence by normative beliefs emanating from secondary sources like media on the intention to adopt smart metering technology. As most studies depicted a positive influence, we concluded it as having strong support.

There was no significant number of studies in the referred articles for hedonic use of IT in the personal green IT context. Therein, we could not analyse the overall impact of SDT constructs and the hedonic use of green IT. However, we identified strong support for intrinsic motivation and moderate support for introjected and external regulations, showcasing a positive influence (Chen et al. 2016; Du et al. 2020; Ke et al. 2019 Lokuge et al. 2020; Lokuge et al. 2021). Other related studies are listed in Table 2.

## 4.3 Summary of the Results

Tables 1 and 2 and Figures 1 and 2 summarise empirical support for the links between the SDT dimensions and personal green IT use. To summarise the empirical results across all investigations, we rated the level of support for each influence as strong, moderate, or mixed, using the criteria used by Petter et al (2008). We allocated 1.0 points to studies with significant, positive results; 0.5 points to studies with significant, positive, and non-significant outcomes (i.e., mixed results); and 0.0 points to studies with a non-significant (Ns) association between the constructs. Furthermore, some investigations revealed strong negative correlations. We deducted 1.0 point from such studies. Then we divided point sum by total number of studies. We looked at the percentage distribution and assigned strong, moderate, or mixed values. 'Strong' support was assigned when the percentage of papers with a positive outcome was 86-100%; 'moderate' support was assigned when the percentage of papers with a positive result was 61-85%; and 'mixed' support was assigned when the percentage of papers with a positive result was 25-60%. These percentages are quite conservative. Many of the correlations characterised as moderate and mixed support would indicate a strong association between the constructs when using a more rigors quantitative technique, such as vote counting (Hedges and Olkin, 1980). The goal of our literature evaluation, however, is not to limit the relationships between constructs to a number (or effect size), but rather to recommend areas for future research Moreover, for the current study we have excluded examining the interdependencies between the independent variables (i.e., SDT constructs) (Koo and Chung 2014; Wati and Koo 2012).

| Relationship | Empirical Studies | Study Result | Overall Result | Conclusion |
|---|---|---|---|---|
| Intrinsic→GIT[2] | (Koo et al., 2015; Bukchin and Kerret, 2020; Wunderlich et al., 2012; Wunderlich et al., 2013b; Wunderlich et al., 2013a; Wunderlich et al., 2019; Brauer et al., 2016; Seidler et al., 2020) | + | 8 out of 10 studies found a positive influence | Moderate |
| | (Koo and Chung, 2014; Wati and Koo, 2012) | Ns | | |
| Integrated→GIT | (Koo and Chung, 2014) | Ns | 2 out of 3 studies found a positive influence | Inconclusive |
| | (Wati and Koo, 2012; Bukchin and Kerret, 2020) | + | | |
| Identified→GIT | (Koo and Chung, 2014; Wati and Koo, 2012; Wunderlich et al., 2012; Wunderlich et al., 2013a) | + | 4 out of 4 studies found a positive influence | Strong |
| Introjected→GIT | (Koo and Chung, 2014; Brauer et al., 2016) | Mixed | 2 out of 6 studies found a positive influence | Mixed |
| | (Wati and Koo, 2012) | Ns | | |
| | (Bukchin and Kerret, 2020; Maity et al., 2019) | + | | |
| | (Wunderlich et al., 2013b) | - | | |
| External→GIT | (Koo et al., 2015; Koo and Chung, 2014; Loock et al., 2013; Bukchin and Kerret, 2020; Wunderlich et al., 2012; Seidler et al., 2020; Wunderlich et al., 2013b; Wunderlich et al., 2013a; Kranz and Picot, 2012; Ali et al., 2019; Wunderlich et al., 2019) | + | 11 out of 12 studies found a positive influence | Strong |
| | (Chen et al., 2017) | Ns | | |

*Table 1: Utilitarian perspective of green IT use*

| Relationship | Empirical Studies | Study Result | Overall Result | Conclusion |
|---|---|---|---|---|
| Intrinsic→GIT | (Shevchuk et al., 2019; Du et al., 2020; Ke et al., 2019) | + | 3 out of 3 studies found a positive influence | Strong |

---

[2] GIT refers the green IT behaviors.

| | | | | |
|---|---|---|---|---|
| Integrated→GIT | (Chen et al., 2016) | + | 1 out of 1 study found a positive influence | Inconclusive |
| Identified→GIT | (Chen et al., 2016) | + | 1 out of 1 study found a positive influence | Inconclusive |
| Introjected→GIT | (Chen et al., 2016) (Zhang et al., 2020; Du et al., 2020; Ke et al., 2019) | Mixed + | 3 out of 4 studies found a positive influence | Moderate |
| External→ GIT | (Panzone et al., 2020) (Whittaker et al., 2021; Huang and Zhou, 2020 Chen et al., 2016; Reeves et al., 2015) (Ke et al., 2019; Du et al., 2020) | Mixed + Ns | 4 out of 7 studies found a positive influence | Moderate |

*Table 2: Hedonic perspective of green IT use*

Table 1 and Figure 1 show the analysis of the literature review at a personal utilitarian level, while Table 2 and Figure 2 show it at the personal hedonic level. The tables and figures indicate a scarcity of research examining motivation and green IT behaviour in both groups, but especially in the hedonic level of analysis. According to the review, most research falls under the utilitarian perspective. The meta-view of the existing research provided here allows a researcher to understand the research gaps and strengths for future studies.

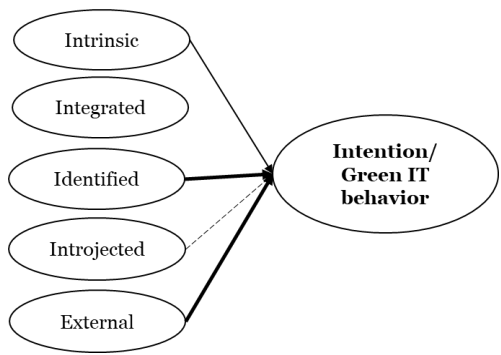

*Figure 1: Conceptual model of strengths of influences from SDT variables on personal green IT behaviour – from a utilitarian value perspective*

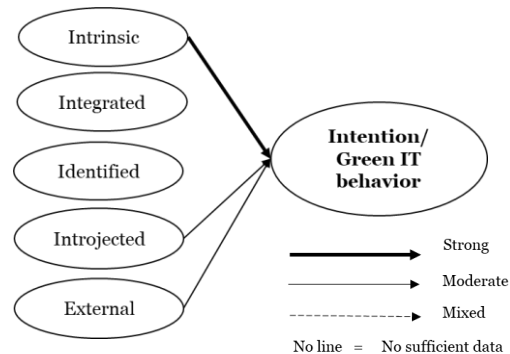

*Figure 2: Conceptual model of strengths of influences from SDT variables on personal green IT behaviour – from a hedonic value perspective*

In applying the analysis of the literature review to the personal hedonic and utilitarian *usage of the internet*, we can identify that identified and external regulations would show the most substantial impact on utilitarian use of the internet. Hence, those two constructs would best manipulate individuals towards responsible personal utilitarian internet use. Apart from that, environmental motivation emanating from pleasure in conserving the environment (intrinsic motivation) could have moderate effects on utilitarian green internet use. An important finding is that, although introjected regulation is vital at the hedonic level, it is not at the utilitarian level. Hence, researchers can focus more on personal norm-related determinants in instilling responsible hedonic internet use behaviours compared to when greening the utilitarian use. However, data are insufficient to make claims on the other relationships.

## 5   Conclusion

This review examined the green IT literature at the personal consumption level on hedonic and utilitarian perspectives of IT use. Although a few studies at the utilitarian level were conducted based on the dimensions of SDT, a sufficient number of studies were not available in the referred databases and journals for the hedonic perspective. Therefore, an overall understanding of the effect of motivation on hedonic green IT use is still to be discovered.

The results provide different operationalizations that were used by past researchers to measure each construct of SDT. As such, this study contributes to knowledge by identifying different ways future researchers can operationalize the SDT constructs. Next, the utilitarian perspective provided varied results in identifying the relationship strengths of green internet use. As per the analysis, identified and external regulations provided the strongest support for green internet use. Intrinsic motivation and introjected regulation-based relationships showed moderate and mixed support, respectively, while data to claim the integrated regulation-based relationship is insufficient. On the other hand, in the hedonic use of IT, we identified that introjected and external regulation-based relationships have moderate support. Furthermore, we identified that different dimensions of five SDT constructs are employed while mainly using the MTE Scale to measure the constructs. As future implications, further studies are required to be carried out to determine the different motivation requirements for utilitarian and especially, hedonic green use of the internet. Based on the current findings, for managing utilitarian and hedonic use, studies can further investigate other green IT behaviours (such as the use of domestic internet of things (IoT) and digital devices), across different geographical territories, gender groups and age groups to identify if current study findings hold true.

# 6　References